\documentclass[aps,pra,preprint,groupedaddress,showpacs,showkeys]{revtex4}
\begin{document}
\title{Geometric aspects of phonon polarization transport}

\author{Mohammad Mehrafarin, Reza Torabi}
\affiliation{Physics Department, Amirkabir University of Technology, Tehran 15914, Iran}
\email{mehrafar@aut.ac.ir}

\begin{abstract}
We study the polarization transport of transverse phonons by adopting a new approach based on the quantum mechanics of spin-orbit interactions. This approach has the advantage of being apt for incorporating fluctuations in the system. The formalism gives rise to Berry effect terms manifested as the Rytov polarization rotation law and the polarization-dependent Hall effect. We derive the distribution of the Rytov rotation angle in the presence of thermal noise and show that the rotation angle is robust against fluctuations.  
\end{abstract}

\pacs{43.35.+d,03.65.Vf}
\keywords{Phonon, Polarization transport, Spin-orbit interaction, Berry effect, Thermal noise}
\maketitle

\section{Introduction}

Berry phase is a non-integrable phase factor arising from the adiabatic transport of a system around a closed path in its parameter space \cite{Berry}. Geometrically, it originates from parallel transport in the presence of a gauge connection in the parameter space \cite{Simon}. This phase factor is amazingly universal and appears in many different contexts \cite{Shapere}. Here, we explore some of its physical manifestations in acoustics.

The analogy between the linearized equations of elasticity and Maxwell equations is well-known (see e.g. \cite {Auld}). One can take advantage of this analogy and predict new phenomena for acoustic waves from their optical counterparts. In particular, polarization phenomena in optics can be mapped onto transverse acoustic waves. For example, the Rytov polarization rotation law \cite{Rytov} and the polarization-dependent Hall effect in optics, which are manifestations of the Berry effect for photons \cite{Tomita,Chiao,Haldane,Berry2,Segert,Liberman,Bliokh,Bliokh2,Onoda,Onoda2,Duval,Sawada}, pertain to transverse acoustic waves as well \cite{Segert,Karal,Bliokh3,Torabi}. 

Acoustical Berry effects in inhomogeneous media were first established \cite{Bliokh3} via the post-geometric acoustics approximation (the acoustical analogue of the post-geometric optics \cite{Bliokh,Bliokh2}).  We have recently treated such media, and in particular those with periodic density variations (phononic crystals), via another semiclassical approach (the wave packet approximation) \cite{Torabi}. A complication that arises in acoustical polarization transport in inhomogeneous media is the coupling between the transverse and longitudinal waves caused by the spatial variations of density and/or elastic coefficients. This means that polarization cannot be maintained. However, the coupling vanishes in the adiabatic limit of small spatial variations which renders polarization an adiabatic invariant \cite{Bliokh3,Torabi}.  In the present study, we introduce yet another approach that is based on the quantum mechanics of spin-orbit interactions. This approach has the advantage of being apt for incorporating fluctuations in the system. The formalism gives rise to a Berry phase describing the rotation of the polarization vector (the Rytov law), and a Berry curvature in the semiclassical equations of motion deflecting the phonons depending on their polarization (the polarization dependent Hall effect). These, of course, reproduce the results of \cite{Bliokh3} (and \cite{Torabi}, as far as non-periodic inhomogeneous media are concerned). We analyze the effect of classical noise on the Rytov law and derive the distribution of the Rytov rotation angle in the presence of thermal noise. It is found that the latter is robust against ubiquitous thermal fluctuations in the system.

\section{The quantum mechanical approach}

The dynamics of the displacement vector filed ${\bf U}({\bf x},t)$ in an elastic medium is governed by (see e.g. \cite{Landau})
$$
\rho \partial_t^2 {U_i} =
\partial_j \sigma _{ij}  
$$
(summation convention implied) where $\rho$ is the mass density of the medium and
$\sigma_{ij}$ is the stress tensor. In an isotropic medium the latter takes the form
$$
\sigma_{ij} =\lambda \delta _{ij}\ \partial_k U_k +\mu (\partial_i U_j+\partial_j U_i)
$$
$\lambda $ and $\mu $ being the Lam\'{e} coefficients. We consider
elastic waves in an isotropic medium whose
density and/or Lam\'{e} coefficients vary with position. By decomposing ${\bf U}$ into a solenoidal (transverse) and an irrotational (longitudinal) part, one can show \cite{Landau} that the shear transverse and the compression longitudinal  modes propagate independently in a strictly homogeneous medium. Introducing sufficiently small (adiabatic) spatial variations, the modes can still be regarded as independent to arbitrary accuracy \cite{Bliokh3,Torabi}. The adiabatic approximation eliminates the coupling between the transverse and longitudinal modes and renders polarization an adiabatic invariant.  

The transverse mode constitutes a two-(polarization) state system. Each state is involved in a double degeneracy in the absence of inhomogeneity. The double degeneracy is the polarization degeneracy;
transverse waves with different polarizations have the
same dispersion in a homogeneous isotropic medium \cite{Kravtsov}. In inhomogeneous media, the refractive-index gradient lifts this polarization degeneracy by coupling the polarization and the translational degrees of freedom \cite{Bliokh3}. The coupling also affects the direction of propagation of the phonon, which is given by its momentum ${\bf p}$. Such spin-orbit interactions also play significant role in the spin transport of photons and electrons 
\cite{Bliokh,Bliokh2,Bliokh4,Berard,Muthur}. The ensuing Berry effects are simply manifestations of this interaction. This motivates an alternative approach to the problem that relies only on the quantum mechanics of the spin-orbit interaction. Thus, following Berry \cite{Berry}, we can derive the geometric phase: The parameter space for adiabatic excursions, here, is the momentum space. The spin-orbit Hamiltonian can be represented by a $(2 \times 2)$ Hermitian matrix coupling the two polarization states. Let us take the point in the parameter space at which the sates are degenerate, as the origin. With reference to this point, the Hamiltonian can be expanded to first order in ${\bf p}$. The most general such matrix satisfying the given conditions depends on the components of ${\bf p}$, and by linear transformation in the momentum space can be brought to the form $\sigma_i p_i$, where $\sigma_i$ are the Pauli matrices. The two helicity eigenstates $|\sigma,{\bf p} \rangle $, with eigenvalues $\sigma p$ ($\sigma=\pm 1, p\equiv|{\bf p}|$), form the non-degenerate normal modes, exactly as for photons \cite{Bliokh2,Bliokh4}. These intersect conically at the origin (the degeneracy point). The Berry curvature is calculated \cite{Berry} to be
$$\nabla_{\bf p} \times \left\langle \sigma,{\bf p} |\ i \nabla_{\bf p}| \sigma,{\bf p} \right\rangle
=-\sigma p^{-3}{\bf p}$$
which is the field of a magnetic monopole of charge $-\sigma$ situated at the origin of the momentum space. It corresponds (in appropriate gauge) to the Berry connection $$\left\langle \sigma,{\bf p} |\ i \nabla_{\bf p}| \sigma,{\bf p} \right\rangle =\sigma {\bf A}_\bot$$
where ${\bf A}_\bot ({\bf p})=\frac{p_3}{p(p_1^2 +p_2^2)}(-p_2,p_1,0)$. The resulting geometric Berry phase $\sigma \int_C d{\bf p} \cdot {\bf A}_\bot$, where $C$ is the phonon trajectory in momentum space, is of opposite signs for the two polarizations. Therefore, for a linearly polarized wave, this Berry phase leads to the rotation of the polarization plane through the angle 
\begin{equation}
\gamma=\int_C {\bf A}_\bot\cdot d{\bf p}=\int_C \cos\theta\ d\varphi \label{1}
\end{equation}
$\theta (\varphi)$ being the zenith (azimuth) angle in the spherical polar coordinates of the momentum space. This is the Rytov law for transverse phonons.

Because the spin-orbit interaction introduces a gauge potential (Berry connection) in the momentum space, the position operator of a transverse phonon of helicity $\sigma \hbar$ acquires an anomalous contribution according to ${\bf x}\rightarrow {\bf r}={\bf x}-\sigma \hbar {\bf A}_\bot$ \cite{Berard}. This is in perfect analogy with the electromagnetic interaction, with the role of the position and momentum interchanged and helicity replaced by charge, of course. (In the post geometric optics/acoustics approximation, the anomalous contribution stems from the representation in which the wave equation is diagonal \cite{Bliokh,Bliokh2,Bliokh3}.)  The physical (observable) position coordinates are no longer the canonical coordinates ${\bf x}=i\hbar\nabla_{\bf p}$, but are now ${\bf r}=\hbar(i\nabla_{\bf p}-\sigma {\bf A}_\bot)$, which are non-commutative: 
$$
[r_i,r_j]=i\sigma \hbar^2 \varepsilon_{ijk} \frac{p_k}{p^3}. 
$$
While $[p_i,p_j]=0$ and $[r_i,p_j]=i\hbar \delta_{ij}$, this provides an example of non-commutative quantum mechanics. Using these commutation relations, the Heisenberg equations 
$$
\dot {\bf p}=\frac{1}{i\hbar}[{\bf p},H], \ \ \ \ \dot {\bf r}=\frac{1}{i\hbar}[{\bf r},H]
$$
yield the semiclassical equations of motion 
\begin{equation}
\dot {\bf p}=-\nabla_{\bf r} H,  \ \ \ \ \dot {\bf r}=\nabla_{\bf p} H+\sigma \hbar \frac{{\bf p}\times \dot {\bf p}}{p^3} \label{2}  
\end{equation}
to first order in $\hbar$. These, of course, reduce to the standard ray equations of geometric acoustics in the classical limit $\hbar \rightarrow0$. In this limit, helicity vanishes and the right/left circularly polarized waves follow the same trajectory. However, according to (\ref{2}), these rays now split due to the effect of the Berry curvature of the momentum space. The deflections from their classical (geometric acoustic) trajectories are given by
\begin{equation}
 \delta {\bf r}=\sigma \hbar \int_C {\frac{{\bf p}\times d{\bf p}}{p^3}}. \label{3}
\end{equation}
The resulting displacements are, therefore, locally orthogonal to the directions of motion of the phonons. This, which is a general feature of spin transport \cite{Onoda,Bliokh4,Horvathy,Murakami,Zhou,Culcer,Bliokh5}, constitutes the polarization-dependent Hall effect of transverse phonons. 
In the present formulation, the effect is a direct consequence of the non-commutativity of the physical coordinates resulting from the spin-orbit interaction. 

The above findings for Berry effects, equations (\ref{1}) and (\ref{3}), coincide with the results of \cite{Bliokh3,Torabi}, which were obtained differently.

\section{Rytov law in the presence of classical noise}

Fluctuations in the system may directly affect the refractive index by randomly perturbing the density and/or the elastic coefficients of the medium. An ubiquitous example is provided by thermal fluctuations. As such, they cause the direction of phonon propagation to fluctuate, so that
$$
{\bf p}(t)={\bf p}_0 (t)+{\bf N}(t)
$$
where suffix 0 indicates the absence of noise and ${\bf N}(t)$ is a noise term- a random process with zero average and small amplitude compared to ${\bf p}_0$. Then, the spin-orbit Hamiltonian $\sigma_i p_i$ still leads to the result (\ref{1}), but the trajectory $C$ is now a fluctuating trajectory- it fluctuates about the noiseless evolution trajectory (assumed cyclic with period $T$). Thus, using (\ref{1}), the resulting change in the Rytov rotation angle $\gamma$ during time $T$ is given by 
\begin{equation}
\Delta \gamma(T) = \frac{2\pi }{T}\int_0^T (\cos \theta- \cos \theta_0) dt \label{4}
\end{equation}
In case, by virtue of the random noise, ${\bf p}$ does not return to its original direction, a non-cyclic contribution also appears. This term must be removed according to the definition of Berry phase for non-cyclic evolution \cite{Samuel}, so that the above result still holds \cite{Chiara}. Writing the integrand in (\ref{4}) as
$$
\cos \theta- \cos \theta_0=\frac{p_z}{p}-\frac{p_{0z}}{p_0} 
$$
and expanding $p$ in terms of $p_0$ to first order in the noise, yields the result
\begin{equation}
\Delta \gamma(T)= \frac{2\pi }{T}\int_0^T (\frac{N_z}{p_0}-\frac{p_{0z}}{p_0^3}\ {\bf p_0}\cdot {\bf N}) dt. \label{5}
\end{equation}
This is the Rytov law in the presence of noise. Thus, $\langle\Delta \gamma\rangle=0$, which means that the average value of the Rytov angle coincides with its noiseless value. To characterize the probability distribution of $\Delta \gamma$ further, we need a definite model for the noise. The uncorrelated noise defined by 
$$
\left\langle {N_i (t)} \right\rangle =0, \quad
\left\langle {N_i(t) N_j(t^\prime)} \right\rangle =2D \delta(t-t^\prime) \delta_{ij}
$$
where angular bracket denotes ensemble average, serves suitably in view of the physical nature of thermal fluctuations. Then, (\ref{5}) yields
$$
\langle \Delta \gamma^2(T) \rangle=\frac{8\pi^2 D}{T} \overline{\left(\frac{\sin \theta_0}{p_0}\right)^2} \propto \frac{1}{T}
$$
for the variance of the distribution, where bar denotes average over the evolution time. The effect of fluctuations, thus, diminishes as $1/T$ so that $\gamma$ coincides with its noiseless value in the adiabatic limit $T\rightarrow \infty$. In other words, the Rytov rotation angle is robust against ubiquitous thermal fluctuations in the system.

\end{document}